\documentclass{ws-ijmpa}
\usepackage[super,compress]{cite}
\usepackage{graphicx}
\usepackage[breaklinks=true]{hyperref}

\begin{document}
\markboth{F. Barreiro Megino {\it et al.},}{Operational Experience and R\&D results using the Google Cloud for High Energy Physics in the ATLAS experiment}

%
\catchline{}{}{}{}{}
%

\title{Operational Experience and R\&D results using the Google Cloud for High Energy Physics in the ATLAS experiment}

\author{Fernando Barreiro Megino} 
\address{University of Texas at Arlington, 502 Yates Street, Arlinton, Texas 76019, United States\\
fernando.harald.barreiro.megino [at] cern.ch}

\author{Kaushik De}
\address{University of Texas at Arlington, 502 Yates Street, Arlinton, Texas 76019, United States\\
kaushik@uta.edu}

\author{Johannes Elmsheuser}
\address{Brookhaven National Laboratory, P.O. Box 5000, Upton, NY 11973-5000, USA\\
jelmsheuser@bnl.gov}

\author{Alexei Klimentov}
\address{Brookhaven National Laboratory, P.O. Box 5000,  Upton, NY 11973-5000, USA\\
aak@bnl.gov}

\author{Mario Lassnig}
\address{European Organization for Nuclear Research (CERN), 1211 Geneva 23, Switzerland\\
mario.lassnig@cern.ch}

\author{Miles Euell}
\address{Google, 655 New York Ave NW, Washington, DC 20001, United States\\
mileseu@google.com}

\author{Nikolai Hartmann}
\address{Ludwig-Maximilians-Universit\"at M\"unchen, Am Coulombwall 1, 
85748 Garching, Germany\\
nikolai.hartmann@physik.uni-muenchen.de}

\author{Tadashi Maeno}
\address{Brookhaven National Laboratory, P.O. Box 5000, Upton, NY 11973-5000, USA\\
tmaeno@bnl.gov}

\author{Verena Martinez Outschoorn} 
\address{University of Massachusetts at Amherst, Amherst, MA 01003, USA\\
Verena.Martinez@cern.ch}

\author{Jay Ajitbhai Sandesara}
\address{University of Massachusetts at Amherst, Amherst, MA 01003, USA\\
jsandesara@umass.edu}

\author{Dustin Sell}
\address{Google, 655 New York Ave NW, Washington, DC 20001, United States\\
dustinsell@google.com}

\maketitle

\begin{history}
\received{Day Month Year}
\revised{Day Month Year}
\end{history}

\begin{abstract}
The ATLAS experiment at CERN relies on a worldwide distributed computing Grid infrastructure to support its physics program at the Large Hadron Collider. ATLAS has integrated cloud computing resources to complement its Grid infrastructure and conducted an R\&D program on Google Cloud Platform. These initiatives leverage key features of commercial cloud providers: lightweight configuration and operation, elasticity and availability of diverse infrastructure.
This paper examines the seamless integration of cloud computing services as a conventional Grid site within the ATLAS workflow management and data management systems, while also offering new setups for interactive, parallel analysis. It underscores pivotal results that enhance the on-site computing model and outlines several R\&D projects that have benefited from large-scale, elastic resource provisioning models. Furthermore, this study discusses the impact of cloud-enabled R\&D projects in three domains: accelerators and AI/ML, ARM CPUs and columnar data analysis techniques. 

\end{abstract}

\ccode{PACS numbers:}

\section {Introduction}
Distributed computing has been the computational pillar of High Energy Physics (HEP), enabling results such as the breakthrough discovery of the Higgs boson~\cite{bib:ATLAS:2012yve, bib:CMS:2012qbp}. The ATLAS experiment\cite{bib:atlas} at the Large Hadron Collider (LHC) uses hundreds of computing clusters distributed globally to run millions of jobs daily while supporting thousands of physicists analyzing more than half an exabyte of data distributed worldwide. These resources form the Worldwide LHC Computing Grid (WLCG)~\cite{bib:wlcg}. If no action is taken, the large computing demands of the LHC are expected to increase by an order of magnitude in the High-Luminosity LHC (HL-LHC) era due to a 5-7x increase in luminosity and a 4-5x increase in event size due to new detectors and higher event rates. A major R\&D effort is under way to close the gap between future needs and expected resources~\cite{bib:atlashllhc}. This paper describes the authors' experience using cloud resources for the ATLAS experiment, both as resembling a standard WLCG Tier-2 site and to investigate capabilities not widely available on WLCG resources or university facilities. Commercial cloud services and infrastructure can enable new ideas and evolution by providing access to architectures and services not available on-site. They can provide complementary sources for computing power, particularly in case of elastic usage, and can significantly reduce maintenance costs.  Similarly the deployment of non-commercial open source cloud technologies at WLCG Grid sites can help in these areas.

US ATLAS initiated a long-term project with Google in 2017 focused on the ``Data Ocean''~\cite{bib:DataOcean} project to demonstrate how storage and compute resources from Google could be seamlessly integrated with the ATLAS distributed computing environment to address the HL-LHC computing challenges starting in 2029. The next phase of the collaboration~\cite{bib:seamless} was launched in 2020 with two years of financial support from the US Department of Energy (DOE).

The success of this  Google project triggered interest from the ATLAS international collaboration, and a 15-month ATLAS-Google project - on which the study is centered - was started in May 2022~\cite{bib:CHEP2023}. The project phase 
was supported through a novel subscription agreement consumption model with Google. The ``Subscription Agreement for US Public Sector'' is negotiated on a case by case basis. In the present case it defined a fixed monthly cost for the usage of resources comparable to a standard WLCG Tier-2 site. The resource usage targets were defined as an average over the subscription period, allowing for dynamic resource consumption depending on the ongoing activities. 

The primary objectives of the phase were: evaluating the feasibility of running a complete ATLAS computing site in Google Cloud; evaluating new architectures such as ARM~\cite{bib:arm} and GPUs for ATLAS workflows; and determining the Total Cost of Ownership (TCO) of such a cloud facility. This paper will focus on the technical aspects of the integration and will describe the most relevant R\&D topics and results. The TCO study results will be discussed elsewhere~\cite{bib:tco}.

The paper is organized as follows.
  Section \ref{adc-overview} provides a high-level overview of the ATLAS distributed computing system, focusing on the elements that will be necessary for the understanding of later sections.
  Section \ref{atlas-cloud-site} illustrates in detail the technical choices and operational experience in running an ATLAS Grid site in the Cloud. 
  Section \ref{rnds} details three R\&D projects utilizing advanced technologies, which are not readily accessible in large quantities at traditional ATLAS Grid sites, and which effectively benefit from the resource elasticity of cloud providers like Google:
\begin{itemize}
  \item GPUs and significant memory capacities for neural simulation-based inference analyses,
  \item ARM CPUs for benchmarking the processors, porting the ATLAS software and validating the results,
  \item Large-scale, interactive analysis clusters to develop columnar analysis methods and compare the performance of different file formats.
\end{itemize}

\section {ATLAS Distributed Computing overview}\label{adc-overview}

The WLCG serves as the primary computational infrastructure for the LHC experiments, including ATLAS. It is a global collaboration of more than 170 computing centers. WLCG provides the computational power and storage capacity needed to handle the exabyte of data generated by the LHC experiments. It is a tiered structure, with the Tier-0 at CERN serving as the primary hub for data storage and initial processing. Subsequent tiers (Tier-1, Tier-2 and Tier-3) are distributed worldwide and provide additional storage and computational resources. 

A Grid site typically consists of a Computing Element (CE)~\cite{bib:ce_arc, bib:ce_htcondor} for workload execution, a corresponding Storage Element (SE) for data storage and the necessary internal and external network bandwidth. These components rely on in-house technologies and middleware developed by the European Grid Infrastructure~\cite{bib:egi}, the Nordic Data Grid Facility~\cite{bib:ce_arc}  and the Open Science Grid~\cite{bib:osg}.

The usage and orchestration of the WLCG resources is defined by the experiment's Computing Model. ATLAS relies on two major usage and orchestration software packages:

\begin{figure}[ht]
\centering
\includegraphics[width=1.0\textwidth,clip]{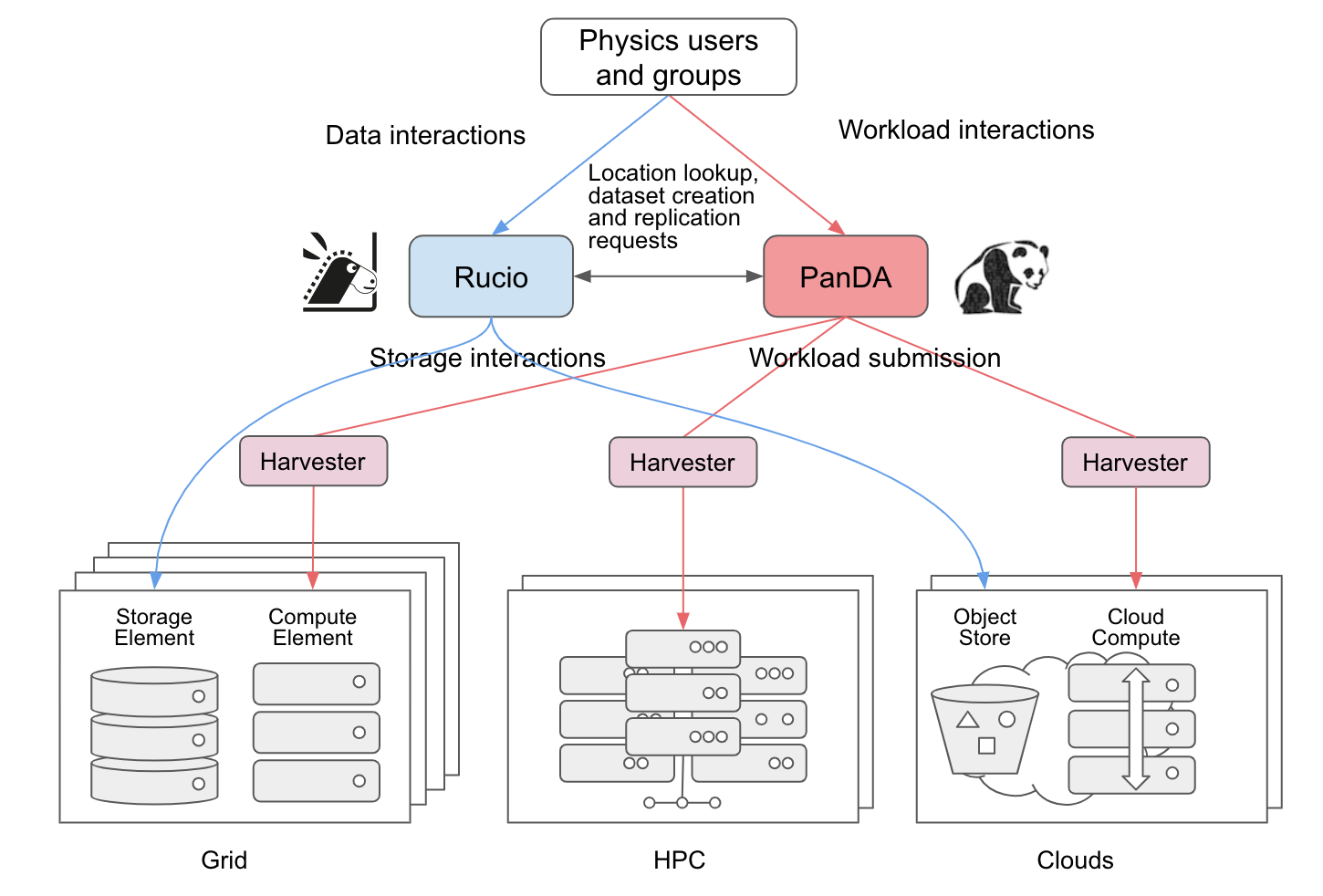}
\caption{Simplified, high-level overview of ATLAS Distributed Computing. Rucio and PanDA serve as the central entry point for ATLAS users and hide the complexity of the underlying distributed resources. ATLAS resources can have different flavors depending of the various Grid initiatives, and have also integrated different types of HPC and cloud resources. 
}
\label{fig:rucio_panda}   
\end{figure}

\paragraph{Rucio}~\cite{bib:rucio} serves as the data management system. It is used by all ATLAS users and systems as an interface to the different SE implementations and protocols. Rucio is aware of the locality of all ATLAS data and interacts with the File Transfer Service (FTS) to handle data replication across the Grid. Rucio manages currently around 700 PB of data on disk and tape for the ATLAS experiment.

\paragraph{Production and Distributed Analysis (PanDA)}~\cite{bib:panda} is the ATLAS workload management system. It serves as the submission point for all organized and individual computational tasks and manages their execution across the Grid. It dynamically matches available computing resources with pending tasks, optimizing for various factors like data locality and computational power. PanDA is able to communicate with different Grid CE implementations and common batch systems, in order to submit and monitor jobs. The communication is done through the resource facing component Harvester~\cite{bib:harvester}, which will also be the integration point for cloud resources as discussed in Section~\ref{compute-integration}. In the case of ATLAS, PanDA manages O(1M) independent jobs per day across a distributed infrastructure of O(1M) concurrent CPU cores.\\

PanDA interacts with Rucio to retrieve data location information, schedule additional replicas of the data, as well as register new data generated by the jobs. Together they manage all ATLAS distributed resources, including non-standard facilities such as High Performance Computing centers (HPC) and cloud resources (see Figure~\ref{fig:rucio_panda}).\\

Another important element in ATLAS and LHC computing model is CVMFS~\cite{bib:cvmfs} (CernVM File System). This read-only file system is used to distribute all of the experiment's software and configuration files across the Grid. It ensures that all computing nodes have consistent and up-to-date software environments, reducing the complexity of managing software dependencies. Software containers are used during the workload execution through PanDA to ensure a consistent operating system setup at all WLCG sites. 

\section {Re-imagining and operating an ATLAS Grid site with cloud technologies}\label{atlas-cloud-site}

\subsection {Considerations when adopting cloud technologies}\label{architecture-considerations}

One of the major benefits of adopting cloud technology is the wide choice of options. Clients can choose to work with low-level infrastructure like virtual machines. In this case, administrators will lift-and-shift\footnote{Lift-and-shift is a strategy for migrating an application from one environment to another without redesigning it. This approach is faster than re-architecting the application, but does not take full advantage of the new platform's capabilities.} standard services (e.g. Storage and Compute Elements) from the source environment to the cloud environment. However, cloud providers also offer high-level, managed services, which reduce considerably the operational load on the service managers. Over time, many of these technologies have become de-facto standards in the IT industry and are available across various cloud providers. Adopting standardized technologies reduces the risk of vendor lock-in and makes it easier for software engineers to leverage existing knowledge and skills. Building on common, standardized technologies often makes it easier to onboard new team members to a project, as there is a larger pool of experts to hire from.

Cloud computing technology enables resource elasticity: the resource consumption is metered and the user pays for the resources actually used, without being bound to a predefined allocation. This elasticity extends to the ability of ``auto-scaling'' resources up or down depending on how much queued work or system load is found. The dynamic resource allocation mechanism helps to optimize costs and adapts transparently to fluctuating workloads.

Some cloud computing services offer different tiers, depending on the quality of service. Cost-saving strategies, such as ``Spot'' compute instances~\cite{bib:gcp-spot}, may also be available. Spot instances offer the same options and performance as the regular instances, but can be preempted with short notice when the cloud provider needs to reclaim the capacity. Spot instances are available at significantly lower prices and are suitable for workloads that can tolerate interruptions, such as repeatable batch jobs.

Cloud providers offer a variety of processor families, accelerators and non-standard machines as part of their resource catalog, and cloud services are optimized to seamlessly incorporate them. The choice of available resources dramatically reduces the time to initiate scientific research, often requiring only days or weeks. In contrast, procuring new types of locally-hosted resources can easily take several months, and these resources are frequently under-utilized until the full adoption in the community has been reached.
Cloud providers offer a variety of processor families, accelerators and non-standard machines as part of their resource catalog, and cloud services are optimized to seamlessly incorporate them. The choice of available resources dramatically reduces the time to initiate scientific research, often requiring only days or weeks. In contrast, procuring new types of locally-hosted resources can easily take several months, and these resources are frequently under-utilized until the full adoption has been reached in the community.

\subsection {Technology Choices}\label{technology-choices}

Based on the considerations described in the previous section, several technology-focused decisions have been made. Instead of installing and managing Grid Storage and Compute Elements on cloud-based facilities, the strategy is to use cloud-native services and standard protocols. This makes it possible to take full advantage of the cloud services' potential, while also reducing the operational effort and cost.

\subsubsection {Storage integration}

Cloud storage has been integrated with Rucio and the FTS middleware~\cite{bib:ruciocloud}. It is based on Object Stores (e.g. Google Cloud Storage and S3) and offers signed URL mechanisms. The Object Store administrator can download a signing key to generate URLs that allow the recipient to access a specific resource for a limited time without needing to log in. 

The signing key can be installed on Rucio and FTS servers, where it is used to generate signed URLs. These are passed on to the users or other systems in order to download, upload or delete files through the standard HTTP protocol. Object Stores are managed by the cloud providers and require no maintenance.

\subsubsection {Compute integration}\label{compute-integration}

\begin{figure}[ht]
\centering
\includegraphics[width=1\textwidth,clip]{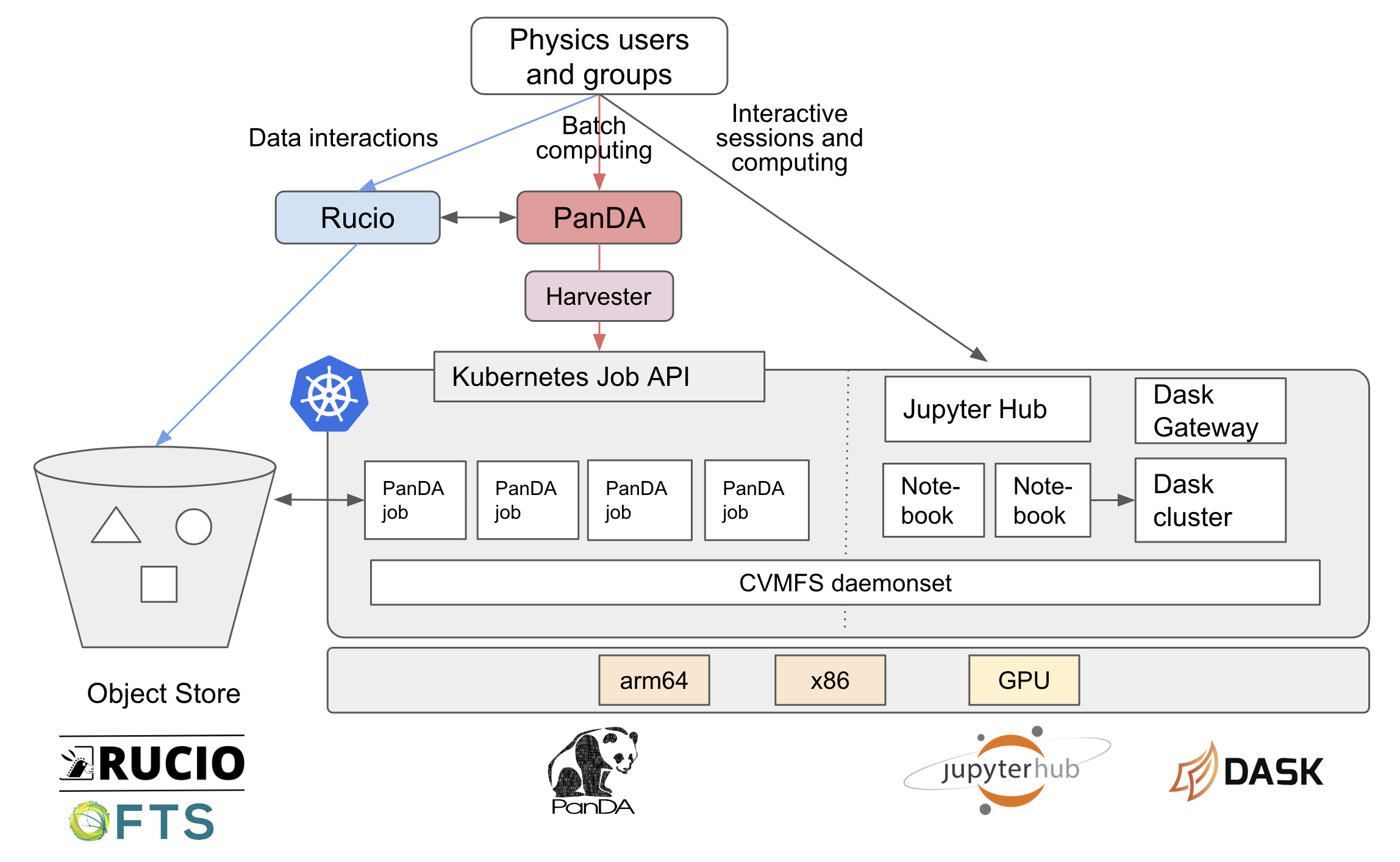}
\caption{Schematic overview of the different components of the site setup. Cloud Object Stores have been integrated with Rucio and FTS, and can be used in the same way as a Grid SE. Batch and interactive computing run as pods in the Kubernetes cluster. PanDA can talk with Kubernetes through Harvester and submit jobs of various sizes and flavors. Users can also connect to JupyterHub, start an interactive Notebook and from there spawn their own Dask cluster for parallel computing. CVMFS is mounted to all Kubernetes pods, delivering the relevant ATLAS software and configurations. The infrastructure layer hosting the Kubernetes cluster consists of nodes with the selected CPU architecture and can optionally include GPUs.}
\label{fig:siteoverview}   
\end{figure}

Figure~\ref{fig:siteoverview} shows the schematic overview of the different components of the ATLAS cloud site. Both batch and interactive workloads are executed as pods in a Kubernetes~\cite{bib:kubernetes} cluster. Kubernetes is offered as a managed service on most cloud providers, e.g. the Google Kubernetes Engine or the Amazon Elastic Kubernetes Service. These managed Kubernetes services offer a complete hands-off experience. The cluster scales up and down within the predefined limits and according to the amount of jobs queued. The cluster is also auto-healing: faulty nodes and pods are replaced by new ones.

Kubernetes provides lightweight, but powerful batch controllers. This choice eliminates the need to run a CE and a batch system. Harvester has been extended~\cite{bib:harvester-kubernetes} to talk natively with Kubernetes for workload submission. CVMFS is installed as a Kubernetes \textit{daemonset} in the cluster, which ensures that CVMFS runs on all nodes and can be mounted into the jobs. Various PanDA queues have been set up to support traditional \textit{x86} jobs, but also to conduct R\&D on GPU and ARM queues. All queues are based on Spot instances and PanDA automatically handles job retries transparently to the end user.

An interactive JupyterHub~\cite{bib:jupyter} data science environment has also been set up on Kubernetes. Jupyter can be integrated with OAuth~\cite{bib:oauth2} providers and the setup has been linked to the ATLAS Identity and Access Management service~\cite{bib:indigoiam}, so that any ATLAS user can connect to the infrastructure. 
This environment has been extended with Dask~\cite{bib:dask, bib:dask-gateway}, so that users can start up private Dask clusters for parallel processing. Installation of Jupyter and Dask is streamlined with Helm~\cite{bib:helm}, Kubernetes' package manager, which offers official charts to deploy both frameworks separately or together.

Software container images have been prepared for data analysis and Machine Learning (ML) applications, which can be selected when starting a Jupyter session (see Section~\ref{sec:pyanalysis}). Jupyter-Dask costs are optimized by distinguishing critical and non-critical node pools, backed by guaranteed and Spot instances, respectively. Important infrastructure components such as web frontends, gateways and schedulers are configured to run on the critical node pool, while workers are configured to run on the non-critical node pool. Dask is resilient to worker failures and will retry the failed jobs. The infrastructure related configuration developed in the course of this project is available in this repository~\cite{bib:gcp4hep}. 

\subsection {Operational experience}\label{operational-experience}

This project has provided significant experience in the operation of an ATLAS site based on cloud-hosted services. Depending on the allocated budget, the site running on Google Cloud has operated with a capacity of either 5,000 or 10,000 vCPUs\footnote{A vCPU is a unit of processing power that represents a single virtualized CPU core. Depending on the cloud provider, this can correspond to a hyper-threaded core.}. The job queue at the ATLAS-Google Cloud site was open to running any ATLAS payload, including production and analysis. To gain a deeper understanding of the behavior of different job types, particularly between CPU and disk-intensive payloads, the queue has been configured to run individual payloads (event generation, simulation, group production, reconstruction, data processing) for periods of several days.

\begin{figure}[ht]
\centering
\includegraphics[width=\textwidth]{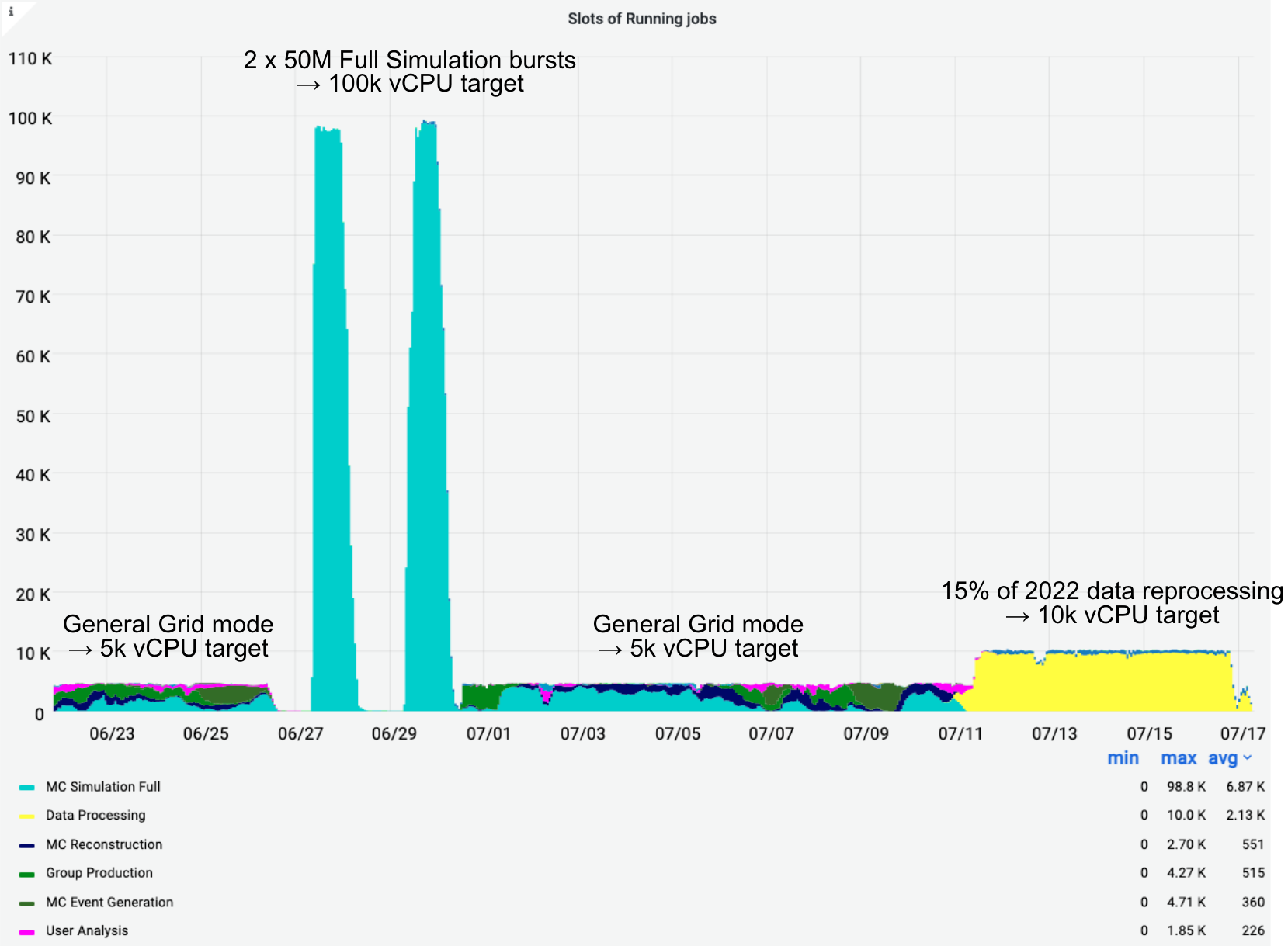}
\caption{Showcase of the flexible and elastic usage ofcloud computing resources: alternating between a generic flat 5k vCPU queue, a 100k vCPU queue to process two Monte Carlo simulation campaigns and data reprocessing on 10k vCPUs}
\label{fig:dynamic}
\end{figure}

The studies have also demonstrated the remarkable elasticity of cloud resources (see Figure ~\ref{fig:dynamic}). Scaling out the site for a brief period to 100,000 vCPUs allowed us to execute two Monte Carlo simulation campaigns with 50 million events in around a day each. This same campaign, when executed on the full Grid, competing with other tasks, took around a week to complete. This elasticity suggests that replicating the setup across multiple regions and cloud providers could potentially provide sufficient computing power to complete all of the experiment's simulation campaigns within days, thus reducing reliance on on-site resources throughout the year. Additionally, 15\% of the 2022 data reprocessing campaign was successfully assigned and executed on 10,000 vCPUs in the Google Cloud within a week.

The queue has been actively utilizing Spot instances. Since the start of operations, the ATLAS-Google Cloud site has experienced a 5\% failure rate in terms of wall-clock time, including all errors related to preemption, system upgrades, software and central infrastructure. This failure rate aligns closely with the failure rate observed at ATLAS pledged resources over the same period.

Queues for research and development activities have also been operated, leveraging special resources such as extremely large memory nodes and GPUs. Further details on these activities are provided in sections \ref{ai} and \ref{arm}.

The operation of this infrastructure, including the set up for multiple R\&D activities and frequent configuration changes has been carried out by full time equivalent fractions of a few workload and data management experts in a true DevOps approach, breaking down traditional boundaries between site administrators, Grid operators and central development teams. This approach has made it possible to adapt and optimize operations effectively in response to evolving needs and challenges during the project.

\subsection {Networking considerations}\label{networking-considerations}

The WLCG relies on a complex network of interconnected Grid sites, utilizing extensive research networks. Each site connects to the LHCONE and LHCOPN~\cite{Martelli:2015mkr} through multiple 10~Gbps links, forming a crucial part of data transfer and sharing.

ATLAS has a dynamic usage pattern for both storage and network resources, using high network bandwidths and throughput. Sites can frequently import or export a significant portion of their storage capacity each month. This flexibility allows for the distribution, balancing and availability of data for processing across various sites.

Major cloud providers like Google, Amazon, and Microsoft operate their own private, global networks. When transferring data to or from cloud data centers, the selected networking tier determines whether data enters or exits the private network at the closest point of presence to the other party. Network usage is clearly metered and billed. Importing data into a cloud data center (ingress) is often free, but there are costs associated with exporting data out of cloud data centers (egress), and also when moving data between different regions or continents. Managing these costs efficiently is essential to reduce networking expenses.

Some ideas being explored to minimize networking costs include:
\paragraph{Task Full Chain} aims to link different stages of data processing within the same location, minimizing the need to export data to other sites for processing at the cost of reducing flexibility. The Full Chain concept has been implemented and technically validated at a small scale on Google.

\paragraph{Network Peering} relies on establishing an interconnect between the cloud network and the LHCONE/LHCOPN networks (not to a single on-site data center). This is a complex project that requires aligning the capabilities of cloud interconnect offerings with a federation of independent sites and where one research network provider needs to act as the intermediary. Additionally, it needs to respect the network and security constraints of the LHCONE/LHCOPN networks, which requires limiting IP ranges to ensure exclusively LHC data is being transferred through the links. Notably, not all peering options offer the same egress rates, so further exploration is needed to determine the preferred option.\\

\vfill

\section {R\&D Projects}\label{rnds}
Several R\&D projects have been pursued with the goal of applying cloud computing infrastructure, features and tools in areas such as ML and analytics for physics. Most projects have taken specific advantage of the resource elasticity to ramp up and down ephemeral compute clusters using e.g. GPUs, ARM CPUs or large amounts of memory, depending on need. This was carried out using PanDA or interactive compute with Jupyter notebooks and Dask task scheduling. The usage of these non-standard resources has been extremely valuable and effective. A new ML-based data analysis technique was developed and the migration of the ATLAS software to ARM CPUs was accelerated. Furthermore compact data formats with columnar data access have been investigated.

\subsection{Accelerators and AI/ML usage}\label{ai}

Modern data analysis, detector simulation, and reconstruction techniques in HEP increasingly rely on ML. This trend is expected to rise even more as the dense collision environments produced during the HL-LHC runs create challenges in the existing trigger and reconstruction software that would require state-of-the-art ML methods to resolve. The common denominator in most applications of machine learning is deep learning and a popular choice are Neural Networks (NN). Training a NN is a parallelizable task for which  GPUs are highly optimized, making them far more efficient than CPUs for training and optimization. Moreover, deep learning workloads can be scaled out across multiple GPUs either in a single node (data parallelism) or across multiple nodes (model parallelism). Frameworks like TensorFlow and PyTorch offer built-in support for multi-GPU training. Having a large scale GPU infrastructure can help accelerate the advancements in the use of ML for ATLAS applications. Currently, only a small percentage of the WLCG sites accessible to ATLAS offer GPU resources, and these resources are typically not homogeneously configured across sites. The elastic GPU infrastructure provided by the Google Cloud simplifies the training of very large ML models. These models can improve the quality of the algorithms currently used and enable new applications which would be otherwise unfeasible.

\subsubsection{Application to neural simulation-based inference analyses in the ATLAS experiment}

Neural Simulation-Based Inference (NSBI) refers to the use of simulated events to approximate probability densities and density ratios that would otherwise be intractable. The challenge of using NSBI in HEP stems from the complexity of the particle detectors like ATLAS. In order to use NSBI in practice, very large ML models are required.

To demonstrate the flexibility of an elastic infrastructure to train large ML models, a new analysis where neural simulation-based inference is used for parameter inference was performed using the ATLAS Google project. The ML model used in this analysis was heavily parallelized to benefit from the elastic infrastructure of the Google Cloud. The model was built from approximately 10,000 individual deep NN.

\begin{figure}[ht]
     \centering
         \includegraphics[width=0.45\textwidth]{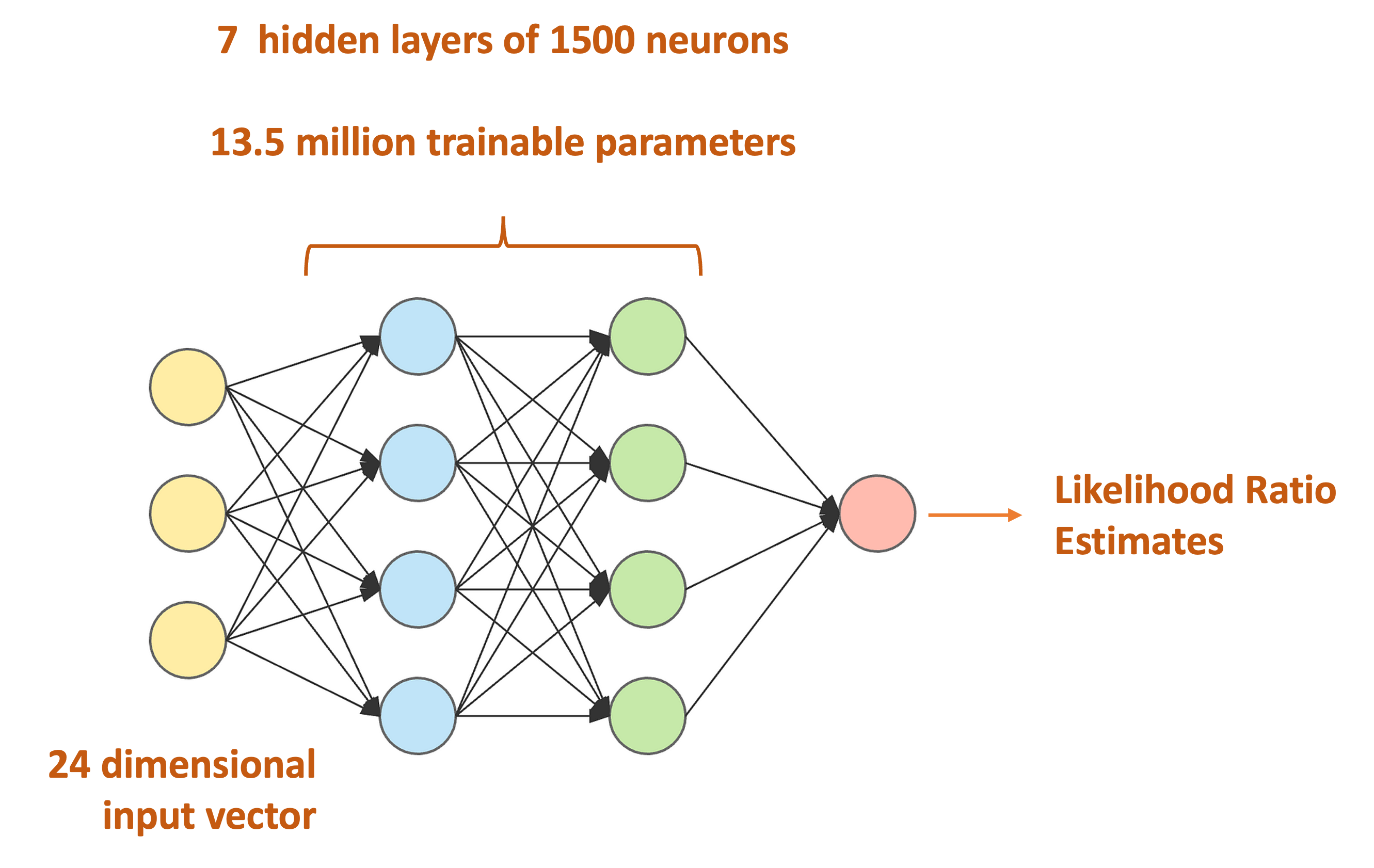}
         \hfill
         \includegraphics[width=0.4\textwidth]{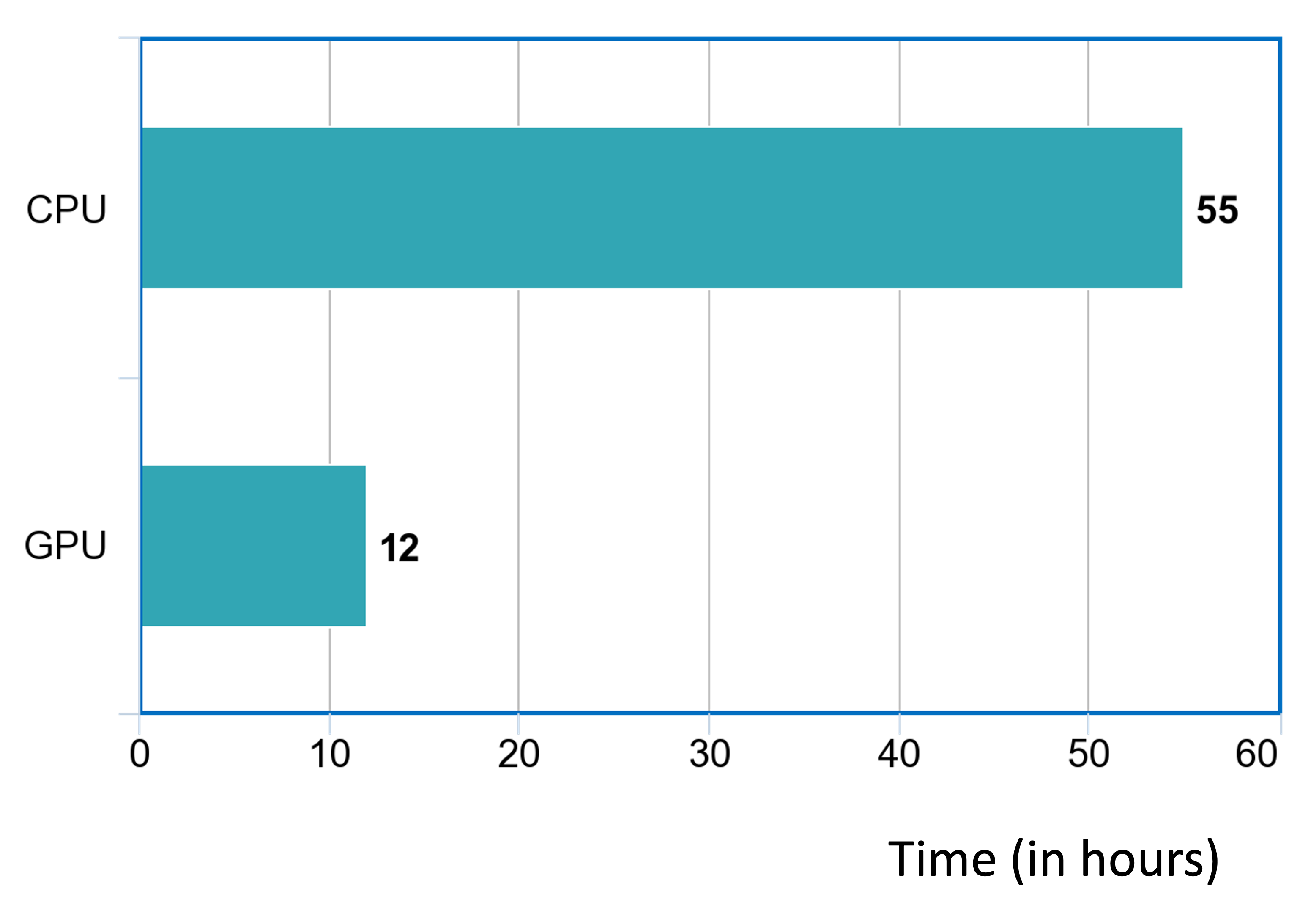}
        \caption{(a) NN architecture of one NN used in the analysis with millions of trainable parameters. (b) Training time comparison, between a CPU and an NVIDIA T4 GPU, for one of the NNs used in the analysis. Training on a GPU offers almost $5\times$ the speed-up compared to the impractically large training time on a CPU, with the same O(1M) simulated events used for each training. This time difference becomes especially relevant when optimizing an ensemble of thousands of NNs.}
   \label{fig:NN_arch}
   \label{fig:GPUvsCPU}
\end{figure}

Each individual deep NN has, itself, a few million trainable parameters. A large-scale GPU infrastructure is necessary for training the models in a timely manner. This is especially true during development, when different models are studied and hyper-parameters are optimized. A comparison of the average time it takes to train a single deep NN with the architecture is shown in Figure~\ref{fig:NN_arch}a, using CPUs at a typical node at a WLCG site or at Google versus using an NVIDIA T4 GPU at the Google site can be seen in Figure~\ref{fig:GPUvsCPU}b.

\begin{figure}[ht]
\centering
\includegraphics[width=\textwidth]{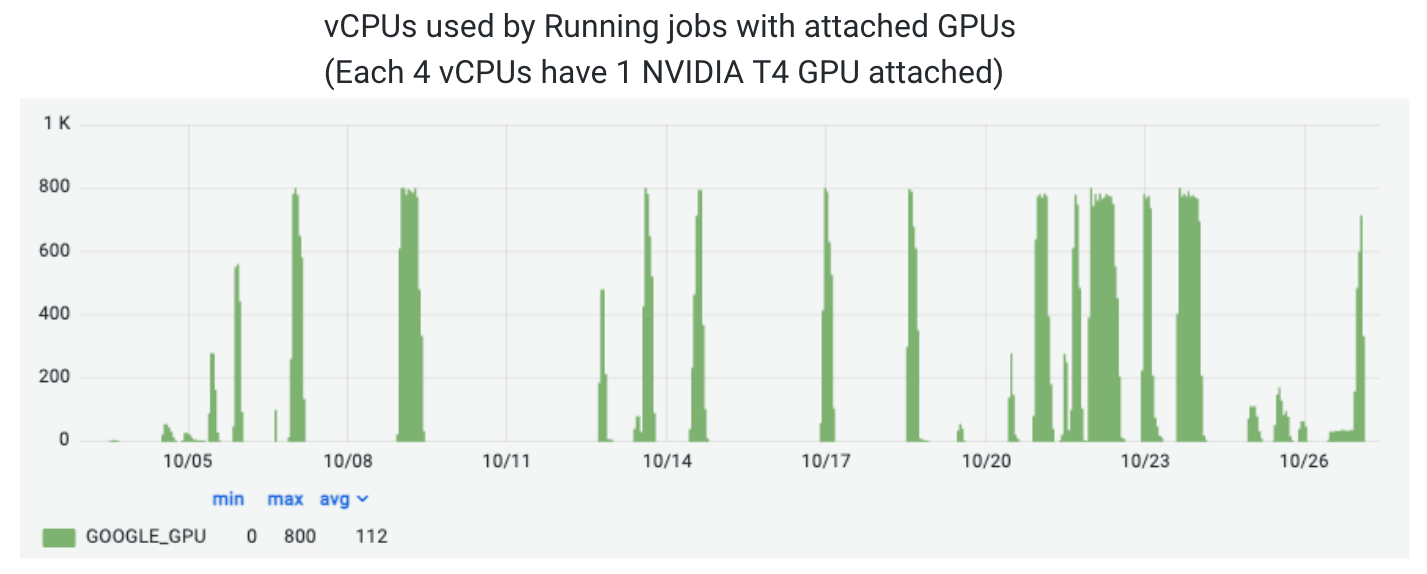}
\caption{On-demand usage on the PanDA Google Cloud GPU queue in October 2022. Each GPU is attached to 4 (until March 2023) or 8 (after March 2023) vCPUs. }
\label{fig:gpu}
\end{figure}

The elastic and rapid on-demand delivery of GPUs using the Google Cloud allowed the successful development of the first NSBI analysis in ATLAS. Figure~\ref{fig:gpu} shows the number of running jobs slots over time on a PanDA queue with up to 200 NVIDIA T4 GPUs. The GPU nodes are provisioned on-demand, ensuring that no infrastructure costs are incurred during the absence of payloads. This PanDA queue currently represents the most active GPU usage in all ATLAS.

\subsubsection{Parameter inference using high-memory cloud CPUs}

Parameter inference at the LHC is performed with a test statistic which is built from profiling, \textit{i.e.} minimizing, the likelihood function with respect to nuisance parameters representing systematic uncertainties. As the analyses at the LHC become more complex, so does the likelihood model used. Current examples of challenging analyses are global Higgs and effective field theory measurements. The ability to start dedicated nodes with very high memory could speed up these calculations considerably.

The use of NSBI methods would increase the complexity of performing parameter inference even more, requiring virtual machines with at least 1~TB of memory to compute and estimate the Hessian matrix of the test statistic. The NSBI analysis performed as part of this project was used to demonstrate how the minimization of the likelihood function $T(\theta)$ and the computation of its Hessian matrix $\nabla^2 T(\theta)$ could be addressed through on-demand high-memory resources on Google Cloud. 

Leveraging auto-differentiation techniques and just-in-time compilation, the problem was parallelized by performing a row-by-row computation of the Hessian matrix. Several high memory nodes, each with 1~TB of memory, were used for the computation~\cite{bib:atlas-google-aiml}. This framework can be extended to other challenging ATLAS analyses and will be necessary for future NSBI approaches.

\begin{figure}[ht]
	\centering
	\includegraphics[width=0.99\textwidth]{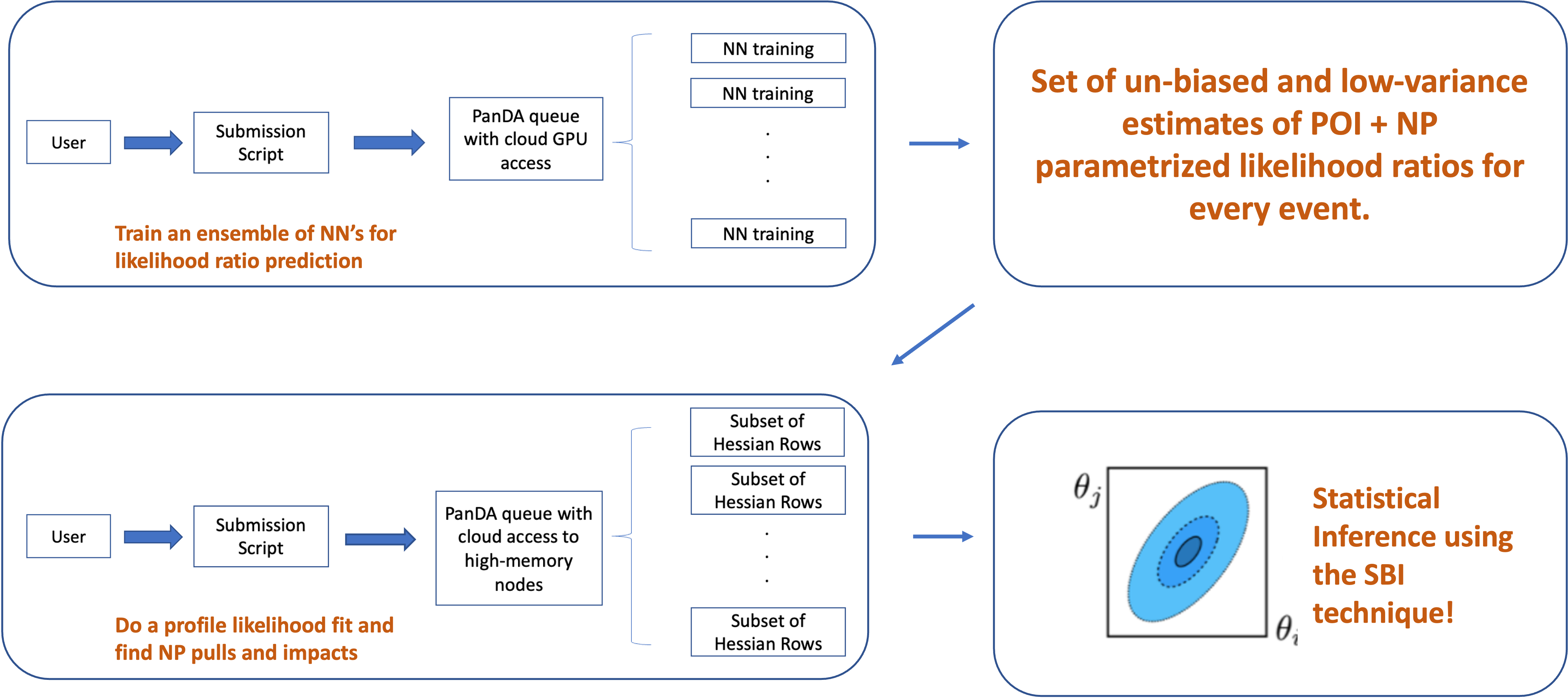}
	
	\caption{Overview of the full neural simulation-based inference (NSBI) analysis workflow using parameters of interest (POI) and nuisance parameters (NP) and how the cloud-based distributed computing infrastructure was leveraged in the different steps.}
	\label{fig:birds_eye_full}
\end{figure}

An overview of the NSBI analysis is shown in Figure~\ref{fig:birds_eye_full}, highlighting the steps making use of the cloud computing infrastructure. The elasticity and power of cloud infrastructure has made the analysis with over a billion NN parameters tractable, reaching a level of precision that would have otherwise been unattainable without engaging in an exceedingly costly infrastructure setup. These developments enable new experimental data analysis directions and open new doors for the use and development of more advanced ML techniques in HEP. The unique resource requirements of this analysis, involving several hundred GPUs with quick turnaround for ML training and nodes with over 1 TB of memory for computations, are unattainable on the ATLAS Grid and pose significant challenges elsewhere, underscoring the efficacy of utilizing cloud resources.

\subsection{Usage of ARM CPUs} \label{arm}
The data processing and simulation software of the ATLAS experiment traditionally uses CPU resources based on \textit{x86} architectures. With an increased dataset obtained during Run 3 and the even larger expected increase of the dataset by more than one order of magnitude for the HL-LHC from 2029 onwards, the ATLAS experiment is reaching the limits of the current CPU resources for data processing and simulation. An extensive program of software upgrades towards the HL-LHC has been set up. The ARM CPU architecture~\cite{bib:atlas-arm} is becoming a competitive and energy efficient alternative. 

Cloud-hosted ARM CPUs provide the opportunity for development and testing of new technologies very easily to prove that the ATLAS software is ready for the HL-LHC data challenge using this type of CPUs. It is possible to validate the port of ATLAS software to ARM CPUs without important capital expenditures for hardware procurement. In the case of a failed validation or other unforeseen circumstances, there would be no unprofitable investment for ATLAS Grid sites.

Using ARM CPUs in the Google Cloud was built on experience gained before in setting up a Grid site on Amazon Web Services (AWS)~\cite{bib:aws} with ARM Graviton2 CPUs. These resources had been integrated into the ATLAS workflow management system PanDA and the ATLAS data management system Rucio in a very similar way as described in the previous section for the Google Cloud.

The full ATLAS Athena~\cite{bib:athena} software stack successfully passed a standard ATLAS physics validation for reconstruction and full Geant4~\cite{GEANT4:2002zbu} detector simulation workflows on ARM. For the Monte Carlo (MC) simulation validation one million events of $t\bar{t}$ decays in the standard ATLAS Run 3 $pp$--collision setup were simulated using the ARM Graviton2 CPUs on AWS. The same events were simulated in parallel on \textit{x86} CPUs using standard WLCG resources. All other steps in the MC event generation, reconstruction and analysis chains were executed on \textit{x86} CPU WLCG resources. The execution on AWS produced 1k files with a total size of $\approx$700 GB and took about 1.5 days on 300 x 8-core vCPU PanDA job slots. 
The comparison showed very good agreement of all physics objects distributions between the events produced with ARM and \textit{x86} hardware, within numerical precision and statistical uncertainties.

Later the reconstruction workflow successfully passed the physics validation on ARM. Similar to the simulation validation, reconstruction workflow samples were produced on ARM Graviton2 CPUs on AWS and compared to regular \textit{x86} CPU samples produced using WLCG sites. All other workflows for MC event generation, simulation and subsequent analysis were executed on \textit{x86} WLCG sites. Thirteen different MC physics processes with 100k events each and no extra pile-up were processed on 130 x 8-core vCPU PanDA job slots and produced 215 GB of output in total. The jobs took about 1.5 days overall.

The Google Cloud Engine Ampere Altra ARM nodes with 48 vCPUs, 192 GB of RAM, 250 GB of storage on a network filesystem and the default Rocky Linux 9 operating system were used for further testing and debugging ARM software builds. The CVMFS file system was used for the distribution of the software. Using this setup, several stable ATLAS offline releases were built and subsequently deployed for production usage through the PanDA workflow management system on WLCG Grid sites with deployed ARM CPUs.

The cloud infrastructure has also provided an opportunity to benchmark a range of resources. The benchmarks currently used, called HEPscore~\cite{Giordano:2020qhu}, are based on experiment workflows. The ATLAS experiment has integrated the following three workflows using Run 3 $pp$--collision workflows using a recent Athena release for \textit{x86} and ARM/aarch64: Sherpa MC event generation of $t\bar{t}$ decays using one CPU thread, data reconstruction using four CPU threads, and Geant4 MC simulation of $t\bar{t}$ decays using four CPU threads.

\begin{figure}[ht]
\centering
\includegraphics[width=0.7\textwidth,clip]{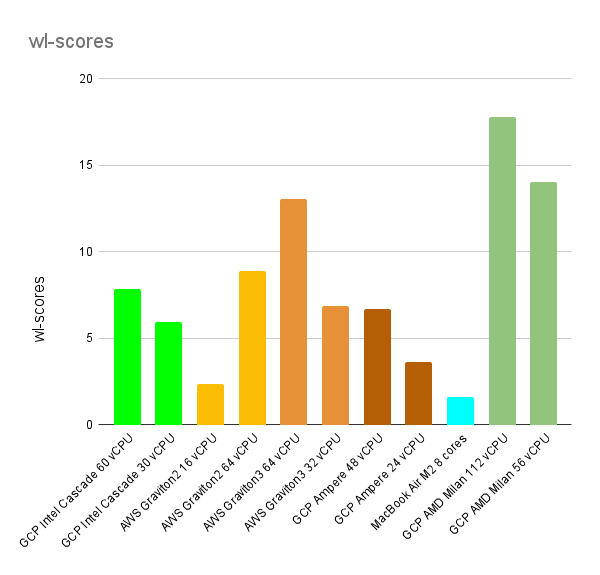}
\caption{HEPscore results for the ATLAS reconstruction workflow and on different \textit{x86} and ARM processors using the full node with all available threads (wl-scores). ``wl-scores'' stands for workload score of the individual HEPScore benchmark. Larger values of wl-scores are better. ARM type CPUs colored brown, orange and blue and are grouped in the middle of both plots as bars 3--9. The \textit{x86} type CPUs are at the very left and right of the plot and colored green.}
\label{fig:HEPscoreA}
\end{figure}

When benchmarking a full CPU node all the workflows are scaled to the number of available CPU cores to fully load the machine. The performance of the ATLAS benchmarks has been tested on several \textit{x86} and ARM machines at Google (Intel Cascade Lake, AMD EPYC Milan, Ampere Altra up to 112 CPU threads), AWS (Graviton2 and 3 with up to 64 threads) and an Apple M2 ARM CPU. Figure~\ref{fig:HEPscoreA} shows the HEPscore results for the ATLAS reconstruction workflow and different \textit{x86} and ARM processors using the full node with all available threads. 
The large differences between the different CPU types can be easily explained with the number of CPU threads available in the different CPU models. The different ARM CPUs show very competitive results in comparison to the Intel and AMD \textit{x86} CPUs tested. 

The successful porting, positive validation and benchmarking results on cloud-hosted CPUs have established a solid foundation for the gradual adoption of ARM CPUs. ATLAS sites are showing interest and starting to include ARM CPUs in their capacity planning.

\subsection{Columnar data analysis}\label{sec:pyanalysis}

The increasing size and complexity of the LHC dataset requires new analysis strategies so that scientific results can continue to be produced in a timely fashion. It is important to both have fast development cycles and efficient processing. Interactive analysis in python, using Jupyter notebooks, provides a way to develop an analysis quickly. In the traditional HEP data analysis model, each collision event is analyzed individually. This model can be inefficient because on the one hand loops over events in interpreted languages like python can be slow and on the other hand each event at the LHC can have different numbers and types of particles. That means even with compiled code there can be inefficiency from CPU cache misses when the data for these different particle types is accessed from non-consecutive memory addresses. A new paradigm, ``columnar data analysis", splits the data processing by column instead of by row. For particle physics applications this then leads to processing by type of particle instead of by collision event. Operations on single columns are run with fast optimized routines, allowing for an analysis workflow that is both interactive and efficient. To be both interactive and scalable it's crucial to have an elastic infrastructure like the one Google Cloud provides.

Analysis workflows in ATLAS entail processing from centrally produced data formats (derived analysis data or DAOD). The new ATLAS \texttt{DAOD\char`_PHYSLITE}~\cite{bib:physlite} data format is designed for rapid analysis. This reduced format already has corrections applied to analysis objects (calibrations) and columns are stored in ``split mode'' in the ROOT~\cite{bib:root} files, meaning efficient reading of single columns is possible. It is therefore an ideal candidate for columnar processing. In addition, this format is planned to be used in ATLAS for the HL-LHC, where the dataset sizes are expected to increase by several orders of magnitude, and highly parallelizable and efficient processing will be critical.  

Dask can be used for scaling interactive workflows to datasets that do not fit in the memory of a single machine or to split up computational workload for parallel processing. Interactive processing requires short feedback loops and therefore the ability to scale the number of workers up and down during a short period of time. The short processing times and the dynamic addition and removal of workers in Dask is also possible using Spot instances. 

The JupyterHub service hosted in the cloud, as detailed in Section~\ref{technology-choices}, provides an ideal infrastructure for this type of physics analysis. When overheads are not considered, the cost of executing a given task on a small cluster over an extended period is comparable to that on a larger cluster within a shorter timeframe, since the amount of CPU-hours will be equivalent. However, the latter scenario significantly enhances user experience and reduces time to physics insights, thereby facilitating rapid iteration and refinement of results by physicists.

An analysis of \texttt{DAOD\char`_PHYSLITE} was carried out on the full ATLAS Run 2 dataset. The ROOT files have a total size of around 100~TB and contain approximately 18 billion recorded and reconstructed collision events. To study different file formats and their throughput from cloud storage, the data was also converted into the Parquet~\cite{bib:parquet} format. In these tests, one Dask worker corresponded to one vCPU, so the terms cores and workers are used interchangeably. One of the main processing steps in a physics analysis using DAOD is data reduction based on the objects of interest for each analysis. Two different workflows were studied for scaling interactive processes with Dask on Google Cloud:

\paragraph{Simple filtering} selects events with two leptons (electrons or muons) to generate a dilepton invariant mass spectrum, or with four leptons to produce a Higgs boson mass peak, without imposing any additional criteria on other event attributes. In these cases, the workflow required reading less than 1\% of the column data of each file and an insignificant amount of computation. 

\paragraph{Object selection} entails more complex criteria, similar to the one described in Ref.~\cite{bib:nikolai-vchep}, that required processing around 10\% of the columns in the dataset.\\

Since data processing can be very fast when using vectorized operations, it is crucial to get performant data delivery from the storage. The  analysis tasks executed exploit the fact that only a fraction of columns (about 10\%) need to be loaded. This is possible since the TTree objects in the ROOT files represent a columnar storage format. Using uproot~\cite{bib:uproot}, the data can be loaded into contiguous arrays in memory for columnar data analysis. However, the data is organized into smaller compression units, called baskets. The data for a single column is split into multiple baskets which are not arranged sequentially in the files. This results in unpredictable access patterns that can lead to slow read speeds due to network latency when accessed remotely. In storage systems currently used for HEP data, this is resolved by using vector reads with the XRootD~\cite{bib:xrootd} protocol or multiple ranges in the header of an HTTP request. With these methods all chunks of binary data for a single column are sent in one response and therefore only a single round trip is required to deliver the data. The HTTP interface to Google Cloud Storage however only supports a single range in the header, requiring the usage of concurrent, single range requests. At the time of these tests, the default implementation in the uproot library used a thread pool with 10 threads. A tenfold increase in loading time was observed when reading 10\% of scattered chunks across a file, compared to downloading the entire file. However, the goal was to continue to explore memory-only workflows that are preferable from the cost efficiency perspective, instead of downloading the whole file, attaching additional fast storage or increasing the memory of the virtual machines which incur additional cost. Therefore, a custom uproot data source was implemented that makes use of the aiohttp~\cite{bib:aiohttp} package which provides concurrent reading based on an event loop instead of a thread pool. This method of asynchronous reading scales better to higher numbers of concurrent requests compared to using threads. The default value in aiohttp for 100 concurrent TCP connections showed the best performance. 
While the initial tests relied on a custom implementation, uproot has more recently incorporated the fsspec interface~\cite{bib:fsspec}, which offers an HTTP implementation utilizing aiohttp.

The Parquet files were processed twice as fast as the ROOT TTree files for the simple filtering and object selection tasks, showing that both tasks are mainly limited by data reading. Data reading in this context refers not only to reading of raw data, but also to reading metadata (structure of the file and available columns), decompression and deserialization. The faster processing of Parquet files can be attributed to 3 main aspects: metadata reading, larger compression units and a data layout better optimized for whole column reading in case of higher dimensional vectors (lists of lists per event). While larger compression units can also be achieved by writing ROOT files with larger basket sizes, the other 2 aspects cannot be improved with different settings using ROOT TTree. The future ROOT RNtuple format is expected to be better in these aspects, but writing of \texttt{DAOD\char`_PHYSLITE} files in RNtuple format and reading with uproot was not implemented yet. Therefore, the larger scale tests performed used the Parquet files.

\begin{figure}[h]
\centering
\includegraphics[width=\textwidth]{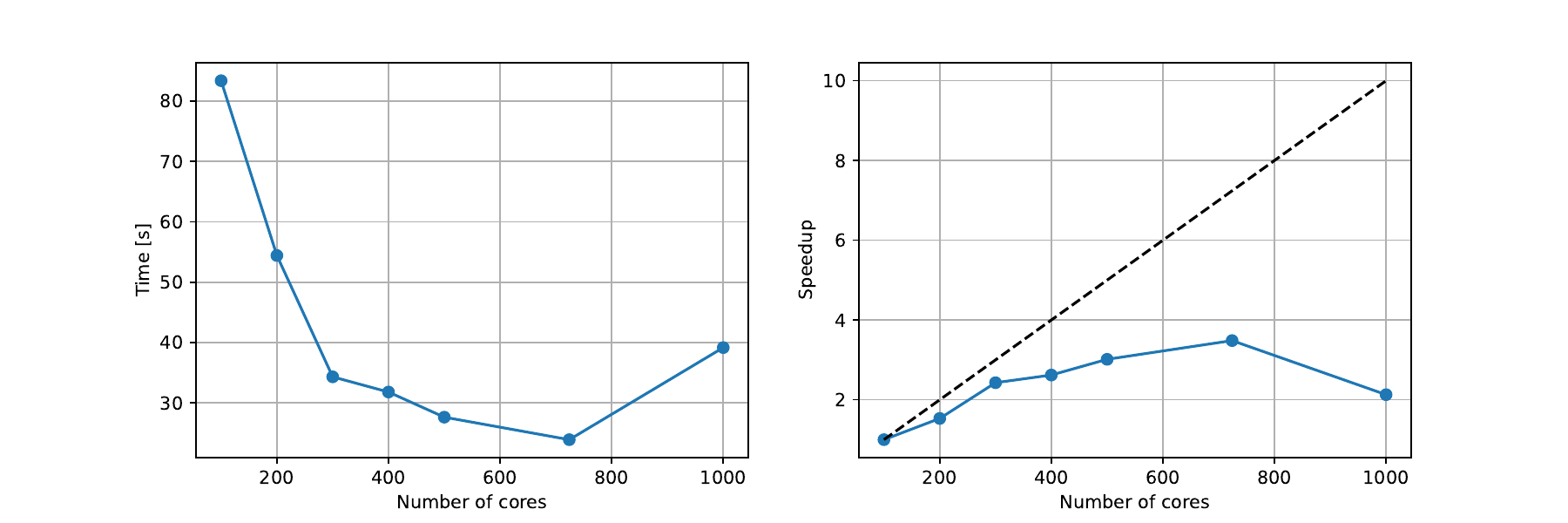}
\caption{Scaling test on Dask clusters processing 10\% of the \texttt{DAOD\char`_PHYSLITE} dataset for the 2-Lepton simple filtering workflow. The {\it Time} in the left plot refers to the total wall-time from starting the first task to finishing the last. The {\it Speedup} on the right plot refers to the ratio between the processing time for the number of cores on the horizontal axis and the measurement with the lowest number of cores (first data point). The black dashed line indicates perfect scaling, e.g. twice the number of cores would lead to a factor of two reduction in processing time.}
\label{fig:dask-scaling-2lepton}
\end{figure}

A systematic scaling test was performed for the simple filtering task using 10\% of all files (see Figure \ref{fig:dask-scaling-2lepton}). The workflow employed Dask's \textit{futures} API, which allows tasks to be submitted to the cluster eagerly, bypassing any prior task graph optimization. One task was defined as a single function processing roughly 100,000 events, corresponding to the row group size in the Parquet files. Almost linear scaling behavior was observed up to 300 cores and no further speedup could be achieved by adding workers beyond 700 cores. This was not ideal behavior, given the simplicity of the task and the lack of communication between workers that could cause sequential bottlenecks. It was possible to increase the number of tasks to several thousand cores by using workloads with longer runtimes (the scaling test used tasks with 2-5s), but this required running the Dask scheduler on a node with larger memory (32 GB). 
For a Dask task stream running the object selection task on 4000 CPU cores the bulk of the processing was finished within 10 minutes.

The nature of large-scale interactive analysis is characterized by its spiky demands, presenting challenges when trying to co-locate it on-site alongside other services, such as the need to preempt batch resources. Currently, Jupyter and Dask setups, are being adopted by the community and are not yet widely implemented across Grid sites. However, the elasticity of cloud services has facilitated rapid progress and provided a significant advantage. This flexibility enabled it to obtain results for each iteration within minutes, allowing for effective debugging and optimization of various software packages tailored for columnar analysis. Furthermore, the performance of ATLAS ROOT files and other formats were compared and it was demonstrated that they can be efficiently processed using cloud storage.

\section {Summary and Conclusions}
Google Cloud computing services have been successfully integrated within the ATLAS workflow and data management systems PanDA and Rucio. State-of-the-art technologies like Kubernetes, non-standard processors like ARM or GPUs and cloud storage are used and have been operated at the scale of a current medium-sized Grid site. Scaling to more than 100k concurrently running vCPUs in PanDA has been demonstrated. The discussed solutions are not ATLAS specific and can be easily adapted by other HEP or non-HEP communities, and may be considered for example in the implementation of analysis facilities for HL-LHC, with all the added values and services offered by the commercial cloud providers. It is planned to continue taking advantage of cloud resources to study novel analysis and data processing techniques. In future work, data exports from cloud data centers to WLCG sites will be optimized, involving research network providers.


\begin{thebibliography}{0}

\bibitem{bib:ATLAS:2012yve}
G.~Aad \textit{et al.} [ATLAS],
{\it Observation of a new particle in the search for the Standard Model Higgs boson with the ATLAS detector at the LHC,}
\href{http://dx.doi.org/10.1016/j.physletb.2012.08.020}{Phys. Lett. B \textbf{716} (2012), 1-29}.

\bibitem{bib:CMS:2012qbp}
S.~Chatrchyan \textit{et al.} [CMS],
{\it Observation of a New Boson at a Mass of 125 GeV with the CMS Experiment at the LHC,}
\href{http://dx.doi.org/10.1016/j.physletb.2012.08.021}{Phys. Lett. B \textbf{716} (2012), 30-61}.

\bibitem{bib:atlas}
  ATLAS Collaboration,
  {\it The ATLAS Experiment at the CERN Large Hadron Collider,}
  \href{http://dx.doi.org/10.1088/1748-0221/3/08/S08003}{JINST {\bf 3} S08003 (2008)}.

\bibitem{bib:wlcg}
Worldwide LHC Computing Grid, URL: \href{http://cern.ch/lcg}{http://cern.ch/lcg} [accessed 2023-11-01]

\bibitem{bib:atlashllhc}
ATLAS Software and Computing HL-LHC Roadmap, \href{https://cds.cern.ch/record/2802918}{CERN-LHCC-2022-005}

\bibitem{bib:DataOcean}
M. Barisits \textit{et al.},
{\it The Data Ocean Project,}
\href{https://doi.org/10.1051/epjconf/201921404020}{EPJ Web of Conferences 214, 04020 (2019)} 

\bibitem{bib:seamless} 
J.~Elmsheuser \textit{et al.}, {\it Seamless integration of commercial Clouds with ATLAS Distributed Computing},
\href{https://doi.org/10.1051/epjconf/202125102005}{EPJ Web Conf. \textbf{251}, 02005 (2021)}

\bibitem{bib:CHEP2023}
F.~H.~Barreiro~Megino \textit{et al.},
{\it Accelerating science : the usage of commercial clouds in {ATLAS} distributed computing. {CHEP2023}, {N}orfolk, {USA}},
{\href{https://indico.jlab.org/event/459/contributions/11636/}{https://indico.jlab.org/event/459/contributions/11636}}

\bibitem{bib:arm}
ARM Architecture Reference Manual for A-profile architecture, URL: \href{https://developer.arm.com/documentation/ddi0487/latest/}{https://developer.arm.com/documentation/ddi0487/latest/} [accessed 2023-08-14]

\bibitem{bib:tco}
ATLAS Collaboration, {\it ATLAS Google Project: Total Cost of Ownership}, to be published in Computing and Software for Big Science

\bibitem{bib:ce_arc}
NorduGrid ARC documents, URL: \href{http://www.nordugrid.org/documents/}{http://www.nordugrid.org/documents/} [accessed 2023-11-01]

\bibitem{bib:ce_htcondor}
HTCondor-CE Overview, URL: \href{https://osg-htc.org/docs/compute-element/htcondor-ce-overview/}{https://osg-htc.org/docs/compute-element/htcondor-ce-overview/} [accessed 2023-11-01]

\bibitem{bib:egi}
About EGI, URL: \href{https://www.egi.eu/about/}{https://www.egi.eu/about/} [accessed 2023-11-01]

\bibitem{bib:osg}
OSG Site documentation, URL: \href{https://osg-htc.org/docs/}{https://osg-htc.org/docs/} [accessed 2023-11-01]

\bibitem{bib:rucio}
M.~Barisits {\it et al.},
  {\it Rucio - Scientific data management},
  \href{http://dx.doi.org/10.1007/s41781-019-0026-3}{Comput.\ Softw.\ Big Sci.\  {\bf 3} (2019) no.1,  11}

\bibitem{bib:panda}
T.~Maeno \textit{et al.}, \textit{PanDA for ATLAS distributed computing in the next decade}, \href{http://doi.org/10.1088/1742-6596/898/5/052002}{J. Phys. Conf. Ser. \textbf{898} (2017) no.5, 052002}

\bibitem{bib:harvester}
T.~Maeno {\it et al.}, {\it Harvester : an edge service harvesting heterogeneous resources for ATLAS},
\href{https://doi.org/10.1051/epjconf/201921403030}{EPJ Web Conf. \textbf{214} (2019), 03030}

\bibitem{bib:cvmfs}
    J.~Blomer \textit{et al.}, \textit{The CernVM File System}, \href{https://doi.org/10.5281/zenodo.4114078}{https://doi.org/10.5281/zenodo.4114078}

\bibitem{bib:gcp-spot}
Google Cloud Platform Spot VMs, URL: \href{https://cloud.google.com/spot-vms}{https://cloud.google.com/spot-vms} [accessed 2023-11-01]

\bibitem{bib:ruciocloud}
M.~Lassnig \textit{et al.},
{\it Extending Rucio with modern cloud storage support: Experiences from ATLAS, SKA and ESCAPE. {CHEP2023}, {N}orfolk, {USA}},
{\href{https://indico.jlab.org/event/459/contributions/11296/}{https://indico.jlab.org/event/459/contributions/11296}}

\bibitem{bib:kubernetes}
Kubernetes, URL: \href{https://kubernetes.io/docs/home/}{https://kubernetes.io/docs/home/} [accessed 2023-06-07]

\bibitem{bib:harvester-kubernetes} 
F.~H.~Barreiro Megino \textit{et al.}, {\it Using Kubernetes as an ATLAS computing site},
\href{https://doi.org/10.1051/epjconf/202024507025}{EPJ Web Conf. \textbf{245} (2020), 07025}

\bibitem{bib:jupyter}
T.~Kluyver \textit{et al.}, \textit{Jupyter Notebooks – a publishing format for reproducible computational workflows}, \href{http://dx.doi.org/10.3233/978-1-61499-649-1-87}{http://dx.doi.org/10.3233/978-1-61499-649-1-87}

\bibitem{bib:oauth2}
OAuth 2.0, URL: \href{https://oauth.net/2/}{https://oauth.net/2/} [accessed 2023-11-01]

\bibitem{bib:indigoiam}
Indigo IAM, URL: \href{https://indigo-iam.github.io/v/current/docs/}{https://indigo-iam.github.io/v/current/docs/} [accessed 2023-11-01]

\bibitem{bib:dask}
M.~Rocklin \textit{et al.}, \textit{Dask: Parallel Computation with Blocked algorithms and Task Scheduling}, \href{http://dx.doi.org/10.25080/Majora-7b98e3ed-013}{http://dx.doi.org/10.25080/Majora-7b98e3ed-013}

\bibitem{bib:dask-gateway}
Dask Gateway, URL: \href{https://gateway.dask.org/}{https://gateway.dask.org/} [accessed 2023-06-07]

\bibitem{bib:helm}
Helm, URL \href{https://helm.sh/}{https://helm.sh/} [accessed 2023-12-04]

\bibitem{bib:gcp4hep}
GCP4HEP repository, URL \href{https://github.com/gcp4hep}{https://github.com/gcp4hep} [accessed 2023-12-04]

\bibitem{Martelli:2015mkr}
E.~Martelli and S.~Stancu,
\textit{LHCOPN and LHCONE: Status and Future Evolution,}
\href{http://dx.doi.org/10.1088/1742-6596/664/5/052025}{J. Phys. Conf. Ser. \textbf{664} (2015) no.5, 052025}

\bibitem{bib:atlas-google-aiml}
  J.~Sandesara {\it et al.},
  {\it ATLAS Data Analysis using a Parallel Workflow on Distributed Cloud-based Services with GPUs,}
  {Proc. CHEP Conf. (2023) }

\bibitem{bib:atlas-arm}
  J.~Elmsheuser {\it et al.},
  {\it The ATLAS experiment software on ARM,}
  {Proc. CHEP Conf. (2023) }

\bibitem{bib:athena}
  P.~Calafiura {\it et al.},
  {\it The athena control framework in production, new developments and lessons learned,}
  \href{http://dx.doi.org/10.5170/CERN-2005-002.456}{Proc. CHEP Conf. (2005)}

\bibitem{GEANT4:2002zbu}
S.~Agostinelli \textit{et al.} [GEANT4],
\textit{GEANT4--a simulation toolkit,}
\href{http://dx.doi.org/10.1016/S0168-9002(03)01368-8}{Nucl. Instrum. Meth. A \textbf{506}, 250-303 (2003)}

\bibitem{bib:aws}
Amazon Web Services Services, URL: \href{https://aws.amazon.com}{https://aws.amazon.com} [accessed 2023-06-09]

\bibitem{Giordano:2020qhu}
D.~Giordano and E.~Santorinaiou,
\textit{Next Generation of HEP CPU benchmarks,}
\href{http://dx.doi.org/10.1088/1742-6596/1525/1/012073}{J. Phys. Conf. Ser. \textbf{1525}, no.1, 012073 (2020)}

\bibitem{bib:physlite}
  J.~Elmsheuser {\it et al.},
  {\it Evolution of the ATLAS analysis model for Run-3 and prospects for HL-LHC},
\href{http://dx.doi.org/10.1051/epjconf/202024506014}{EPJ Web Conf. \textbf{245}, 06014 (2020)}
  
  \href{https://doi.org/10.1051/epjconf/202024506014}{10.1051/epjconf/202024506014}

\bibitem{bib:root}
  R.~Brun {\it et al.},
  {\it ROOT - An Object Oriented Data Analysis Framework},
  \href{https://doi.org/10.1016/S0168-9002(97)00048-X}{Nucl. Inst. \& Meth. in Phys. Res. A 389 (1997) 81-86.}

\bibitem{bib:parquet}
Apache Parquet, URL: \href{https://parquet.apache.org}{https://parquet.apache.org} [accessed 2023-10-18]

\bibitem{bib:nikolai-vchep}
N.~Hartmann \textit{et al.} [ATLAS Software and Computing],
\textit{Columnar data analysis with ATLAS analysis formats,}
\href{http://dx.doi.org/10.1051/epjconf/202125103001}{EPJ Web Conf. \textbf{251}, 03001 (2021)}

\bibitem{bib:uproot}
J. Pivarski {\it et al.}, {\it Uproot}, URL: \href{https://zenodo.org/records/10023419}{https://zenodo.org/records/10023419} [accessed 2023-10-18]

\bibitem{bib:xrootd}
XRootD, URL: \href{https://xrootd.slac.stanford.edu}{https://xrootd.slac.stanford.edu}  [accessed 2023-10-18]

\bibitem{bib:aiohttp}
aiohttp, URL: \href{https://docs.aiohttp.org}{https://docs.aiohttp.org} [accessed 2023-10-18]

\bibitem{bib:fsspec}
fsspec, URL: \href{https://github.com/fsspec/filesystem_spec}{https://github.com/fsspec/filesystem\_spec} [accessed 2023-10-18]

\end{thebibliography}
%
%

\end{document}